\documentclass[11pt]{article}
\usepackage[brazil]{babel}
\usepackage{url}
\usepackage{amssymb,amsxtra}
\usepackage{amsmath,color,graphicx}
\usepackage{lineno}
\usepackage{graphicx}
\RequirePackage{geometry}
\usepackage{indentfirst}
\usepackage[breakable]{tcolorbox}
\usepackage{lineno,hyperref}
\usepackage{alltt}
\usepackage{amsfonts}
\usepackage{amsmath,amssymb}
\usepackage{multirow}
\usepackage{eqparbox}
\usepackage{upquote} 
    \usepackage{eurosym} 
    \usepackage[mathletters]{ucs} 
    \usepackage{fancyvrb} 
    \usepackage{grffile} 
\definecolor{incolor}{HTML}{303F9F}
    \definecolor{outcolor}{HTML}{D84315}
    \definecolor{cellborder}{HTML}{CFCFCF}
\definecolor{cellbackground}{HTML}{F7F7F7}
\definecolor{darkgreen}{rgb}{0.00,0.50,0.00}
\definecolor{purple}{rgb}{0.67,0.13,1.00}
\definecolor{lgray}{rgb}{0.60,0.60,0.60}

\begin{document}

\def\chaptername{}
\def\contentsname{Sum\'{a}rio}
\def\listfigurename{Figuras}
\def\listtablename{Tabelas}
\def\abstractname{Resumo}
\def\appendixname{Ap\^{e}ndice}
\def\refname{\large Refer\^{e}ncias bibliogr\'{a}ficas}
\def\bibname{Bibliografia}
\def\indexname{\'{I}ndice remissivo}
\def\figurename{\small Fig.~}
\def\tablename{\small Tab.~}
\def\pagename{\small Pag.}
\def\seename{veja}
\def\alsoname{veja tamb\'em}
\def\na{-\kern-.4em\raise.8ex\hbox{{\tt \scriptsize a}}\ }
\def\pa{\slash \kern-.5em\raise.1ex\hbox{p}\ }
\def\ro{-\kern-.4em\raise.8ex\hbox{{\tt \scriptsize o}}\ }
\def\no{n$^{\underline{\rm o}}$}

\setcounter{tocdepth}{3}

\clearpage
\pagenumbering{arabic}

\thispagestyle{empty}
\parskip 8pt
\begin{center}
{\huge \bf Is time one-dimensional? }\\
\ \\
\vspace*{0.5cm}
{\Large \bf \it Francisco Caruso\,$^{1}$}\\[2.em]

{{$^{1}$ Centro Brasileiro de Pesquisas F\'{\i}sicas, Coordena\c{c}\~{a}o de F\'{\i}sica de Altas Energias, 22290-180, Rio de Janeiro, RJ, Brasil.}}
\end{center}

\vspace*{0.5cm}

\noindent \section*{Abstract}

In the History of Ideas, a succession of philosophical and scientific achievements, concerning the concept of space and its dimensionality, were essential to contribute, after a long period, to the theoretical possibility of thinking physical time with more than one dimension. Meanwhile, such a progress brought with it the expectation that one can either understand the role of dimensionality in the World or disclose how certain physical phenomena depend on it. Some of these issues are sketched throughout the text, as well as those remarkable moments in the History of Science where important contributions were made in order to give a satisfactory answer to the following inescapable question: -- How many dimensions does time have?

\begin{flushright}
\begin{minipage}{6.5cm}
\baselineskip=10pt
\vspace*{0.5cm}

\small{\textit{It is a pleasure to dedicate this work to Ruben Aldrovandi, knowing that friends will live forever in the multiple dimensions of individual and collective memories.}}
\vspace*{0.5cm}
\end{minipage}
\end{flushright}

\section{Introduction}

In a certain sense, the project aimed to explaining time dimensionality has to do with the attempt to derive the experience of time from perception of brief intervals which is similar to the effort of relating human's knowledge of space to the laws of visual perception~\cite{Friedman}, as briefly reviewed in Sections~\ref{Sec-1} and \ref{Sec-2}. In this way, we are led to believe in common sense, \textit{i.e.}, that space has three dimensions and time has one.

The idea of a fourth space dimension had a significant revival with the studies of several mathematicians in the 16th and 17th centuries. The first speculations of new dimensions in space are mentioned in Section~\ref{first-speculations}, as a result of two tendencies: the dissemination of a geometric description of Nature and advances in the field of Geometry by figuring higher-dimensional spaces.

For completeness, in the sequel, it is mentioned how the problem of space dimensionality has been approached in modern physics, starting from Kant's conjecture (Section~\ref{kant}). Non-Euclidean geometries have also an important influence in the study of higher-dimensional spaces (Section~\ref{non-euclidean}). After this, the fourth dimension began to be seen as a time-like component of the new space-time concept introduced by Einstein in Physics (Section~\ref{spacetime}). In Section~\ref{role}, it is stressed how metric spaces play important role in the modern discussion of space-time dimensionality. A few examples of multidimensional time are sketched out in Section~\ref{roots}, and some concluding remarks are given in Section~\ref{conclusion}.

\section{Prelude to an one-dimensional time}\label{Sec-1}

Perhaps the origin of the unidimensional character of time can be traced back to the abandonment of the cyclical time, diffused in the Classical Greek thought. Such a conception was actually developed in the Stoic theory of the \textit{eternal return} of the same cosmic cycles. From its dawn, Christianity, admitting the beginning of Creation, abandons this cyclical conception and adopts a linear time, whose end would be linked to sublimation in a higher sphere, the \textit{Final Judgment}. More specifically, it was Augustine of Hippo, also known as Saint Augustine, who presented the first philosophical theory of time, based on a specific origin (the crucifixion of Christ) and on the conviction that time is measured by human consciousness of the ``rectilinear'' movement of history, irreversible and unrepeatable \cite{Juliao}.

In his book \textit{On Genesis}, written eight years before the \textit{Confessions}, Augustine views \textit{time} as a creature of God, and so time exists before human consciousness and, in this sense, is \textit{objective}. However, in his \textit{Confessions}, he views time also as a phenomenon of human consciousness, which is clearly a \textit{subjective} account of time \cite{Hernandez}. Indeed, corroborating the second conception, we must remember his consecrated answer to the question: What, then is time? -- ``If no one asks me, I know; if I want to explain to a questioner, I do not know.'' \cite{Augustine}. This dichotomy between objectivity and subjectivity will always permeate studies about time.

Throughout history, time has been treated from different points of view, which can be generically associated to its cosmological, gnosiological and ethical-religious meanings. The last two categories involve concepts considered intrinsic to mankind. They correspond, respectively, to a phenomenological, subjective or idealistic perception of time, opposed to ethical, moral and religious conceptions.

A typical example of the use of perception, intended as an attempt to explain the dimensionality of time, can be found in Whitrow's book \textit{The Natural Philosophy of Time}, where the author devoted three pages to the possibility of time being multidimensional \cite{Whitrow-1}. His brief presentation of the theme revolves around two points, namely: the issue of prior knowledge and the reversal of perspective. The first is linked to the capacity of dreaming, while the second refers to the possibility of seeing a simple drawing of a cube in a moment in one perspective, and in another in a different moment. The conclusion to which he is led is that there is no compelling reason why time should have more than one dimension. In his words \cite{Whitrow-2}:
\newpage 

\begin{quotation}
\noindent\baselineskip=10pt{\small{Moreover, it is difficult to believe that our system of physics, based on the concept of a unidimensional time variable, could be as successful as it is if in fact we inhabit a world in which time had two or more dimensions. Consequently, although it has not been possible to prove the physical time must necessarily be restricted to one dimension, there appears to be no reason so far for doubting that it is}.}
\end{quotation}

Throughout this essay, the focus is narrow on the scientific meaning of time, leaving aside other aspects related to its cosmological significance, such as those involving mythology and philosophy.

\section{More than three space dimensions was not acceptable in the Classical Greek Philosophy}\label{Sec-2}

The first step in the history of the fourth dimension was actually an attempt to deny its existence. Indeed, the impossibility of a fourth dimension was actually sustained by Aristotle of Stagira. In effect, in his \textit{De Caelo}, which consists of Four Books, he treated this impossibility just right in the first paragraph of Book 1, saying, in summary, that~\cite{Barnes}:

\begin{quotation}
\noindent\baselineskip=10pt {\small
\noindent A magnitude if divisible one way is a line, if two ways a surface, and if three a body. Beyond these there is no other magnitude, because the three dimensions are all that there are, and that which is divisible in three directions is divisible in all.}
\end{quotation}

In this same paragraph, the Stagirite continues giving a cosmological jus\-ti\-fi\-cation of this number three by appealing to its divinization, sustained by the Pythagoreans. Quoting him,

\begin{quotation}
\noindent\baselineskip=10pt {\small
\noindent For, as the Pythagoreans say, the universe and all that is in it is determined by the number three, since beginning and middle and end give the number of the universe, and the number they give is a triad. And so, having taken these three from nature as (so to speak) laws of it, we make further use of the number three in the worship of the Gods.}
\end{quotation}

Such kind of identification between the tri-dimensionality of space and God's will is recurrent in the history of science. Johannes Kepler, for example, asseverated that three is exactly the number of dimensions due to the Holly Trinity~\cite{Pauli,Jung}.

The second necessary (but not sufficient) step toward the conception of the fourth dimension has to do with the systematization of the geometric knowledge in Ancient Greece, as detailed in \cite{fourth-dimens}.

Some centuries later, the Greek astronomer Claudius Ptolemy, in his (lost) book \textit{On Distance}, published in 150~a.C., would have giving a ``proof'' about the impossibility of the fourth dimension, based on the very fact that it is impossible to draw a fourth line perpendicular to three mutually perpendicular lines. This is indeed not a proof, but rather reinforce that one is unable to visualize the fourth dimension from which one cannot conclude about its non existence.

To the best of our knowledge, speculations and new ideas about the existence of a fourth dimension had to wait for the middle of the 16th century on to be strengthened, when a more propitious intellectual atmosphere is to be found, as shown in \cite{fourth-dimens}.

For the moment, it is relevant to stress that the two greater synthesis of the Classical Greek Philosophy -- that of Aristotle and that of Euclid -- considered impossible the existence of more than three spatial dimensions.

\section{New speculations about the fourth dimension}\label{first-speculations}

Back to the fourth dimension, the idea of a new spatial dimension was revived by the studies of several mathematicians in the 16th and 17th centuries. Indeed, the Italian physicist, philosopher, mathematician and physician Ge(i)rolamo Cardano and the French mathematician Fran\c{c}ois Vi\`{e}te considered such ``additional'' dimension in their researches on quadratic and cubic equations. The same did the French mathematician and physicist Blaise Pascal in his study named \textit{Trait\'{e} des trilignes rectangles et le leurs onglets}~\cite{Pascal}, when, generalizing his ``trilignes'' from the plane to the space and beyond, he wrote: ``\textit{The fourth dimension is not against the pure Geometry}.''

During the 18th century, the theme of the fourth dimension was treated again from a different perspective, \textit{i.e.}, by associating it to \textit{time} no more to \textit{space}. We are talking about the contribution of the French mathematician Jean le Rond d'Alembert and his proposal in the entry ``Dimension'' wrote for the \textit{Encyclop\'{e}die ou Dictionnaire Raisonn\'{e} des Sciences, des Arts, et de M\'{e}tiers}, published between 1751 and 1772, by Denis Diderot and himself.

Time was considered also as a fourth dimension by the Italian-French mathematician and astronomer Joseph-Louis Lagrange, in his books \textit{M\'{e}canique Analytique}, de 1788, and \textit{Th\'{e}orie des Fonctions Analytiques}, de 1797. Later, Lagrange says something like: One can consider the Mechanics as a Geometry in four dimensions and the Analytical Mechanics as an extension of the Analytical Geometry, developed by Descartes in his book \textit{La G\'{e}om\'{e}trie}, published in 1637~\cite{Kline}.

In the beginning of the 19th century, more specifically in 1827, in the book \textit{Der Barycentrische Calcul}, the German mathematician Augustus Ferdinand M\"{o}bius rejected the existence of the fourth dimension when he observed that geometrical figures cannot be superimposed in three dimensions since they are the mirror images of themselves~\cite{Moebius}. Such a superposition, however, could happen just in a four dimensional space but, ``\textit{since, however, such a space cannot be thought about, the superposition is impossible}''~\cite{Kline}.

The fourth dimension was also proposed by the German physicist and mathematician Julius Pl\"{u}cker in his book entitled \textit{System der Geometrie des Raumes}, published in 1846, in which he affirm that planes are nothing but collections of lines, as the intersection of them results in points. Following this idea, Pl\"{u}cker said that if lines are fundamental elements of space, then space is four-dimensional, because it is necessary four parameters to cover all the space with lines. However, this proposal was rejected because it was saw as metaphysics. But, in any case, it was quite clear for many mathematicians that the three-dimensional Geometry had to be generalized~\cite{Manning}.

It is important to stress that before, in 1748, and later, in 1826, the Swiss physicist and mathematician Leonhard Euler and the French mathematician Augustine Louis Cauchy, respectively, had tried to represent lines in space. In 1843, the English mathematician Arthur Cayley had developed the Analytical Geometry in a $n$-dimensional space, taking the theory of determinants (name due to Cauchy) as a tool. Soon, in 1844, the German mathematician Hermann G\"{u}nter Grassmann published the book \textit{Die Lineale Ausdehnungslehre, ein neuer Zweig der Mathematik}, in which he thought on a $n$-dimensional Geometry, stimulated by the discovery of the quaternion, announced by the Irish mathematician and physicist Sir William Rowan Hamilton, in 1843~\cite{Boyer,Kline}.

Actually, the conjectures about the fourth dimension acquire more soundness from the development of the so-called non-Euclidean Geometries in the 19th century~\cite{Caruso-02}. Let us now summarize how it happened.

\section{Kant's conjecture and beyond}\label{kant}

At this point, it is convenient to go back in time to summarize Kantian original contribution to the problem of space dimensionality, due to its impact on almost all the modern discussion about this issue.

In his precritical period, Kant speculated that space dimensionality could be related to a particular physical law, namely the Newtonian gravitational force. In a nutshell, this fruitful conjecture was settled down in the framework of a philosophical debate concerning the causal explanation of the World, which, at that time, demanded him and other scholars to take a stance on the well-known problem of \textit{body-soul} interaction. From a metaphysical point of view, three possibilities of \textit{causation} where admitted with the aim of clarifying how causes are related to their effects: occasionalism, pre-established harmony and the \textit{influxus physicus}. The young Kant positioned himself in favor of the third option, which means to accept that the interaction among bodies and soul should have a physical cause. He was trying to provide a foundation for the metaphysics of nature, understanding that its task is to discover the inner force of things, the first causes of the law of motion and the ultimate constituents of matter. In order to grasp the Newtonian influence on the young Kant, it is enough to recall what is written in the Preface to the first edition of Newton's \textit{Principia}:

\begin{quotation}
\noindent\baselineskip=10pt {\small
\noindent [...] for the whole burden of philosophy seems to consist in this -- from the phenomena of motions to investigate the forces of nature and then from these forces to demonstrate the other phenomena.}
\end{quotation}

In doing so, the young Kant, in his Ph.D. thesis, advanced a very fruitful conjecture: that a physical law should depend on space dimensionality \cite{Kant}. However, now it is well known that he could not actually prove it \cite{Epistemologia,Kant-Studien}. The result of his analysis refers, actually, to the dimensionality of the \textit{extension}, not \textit{space}.

Albeit the basic idea of somehow relating dimensionality to the gravitation law was abandoned during the critical period of Kantian philosophy~\cite{Britain}, it is still, unquestionably, a milestone in the contemporary discussion of this essential attribute of space. Frustrated for not achieving the expected result in his thesis, Kant, in his critical period, admits the \textit{a priori} nature of space an time. Thereby, their dimensionalities do not have to be explained. A fragment, published in his \textit{Opus Postumum} \cite{Postumum}, reveals that the mature Kant revisited the theme of \textit{dimensionality}, including that of time. Ironically, only the following fragment survived: ``\textit{The quality of space and time, for example, that the first has three dimensions and the second, only one, are principles that} (...)''.

From the physical point of view, a deeper comprehension of Kant's conjecture can only be reached by means of the concept of \textit{field}. Following Paul Ehrenfest, it was through the solution of the Laplace-Poisson equation for planetary motion in $n$-dimensional Euclidean space that one can prove, straightforwardly, the relation between the exponent of the Newtonian potential (through which Newtonian gravitational force is determined) and the dimensionality of space is established.

A systematic scientific investigation inspired on Kant's conjecture began indeed with the seminal contributions of Ehrenfest \cite{Ehrenfest1,Ehrenfest2}. His basic idea to shed light to the problem of space dimensionality was that one must try to identify particular aspects of a physical phenomenon, called by him ``singular aspects'', which could be used to distinguish the Physics in three-dimensional space from that in $D$-dimensions. Formally, what he did to carry on this project was to postulate that the form of a differential equation -- which usually describes a physical law in a three-dimensional space -- is still valid for an arbitrary number of dimensions \cite{Causa}. Therefore, the explanation of space dimensionality is based on a \textit{causa formalis} (the form of the equation), while, in the framework of the original Kant's idea, it elapse from a \textit{causa efficiens} (the gravitational force). Once the general solution of such an equation could be found, he required that it should be a stable solution. As an example, Ehrenfest assumed the motion of a planet under a central force, associated with the Newtonian gravitational potential, to be still described by the Laplace-Poisson equation, keeping the same power of the Laplacian operator $\Delta$ and making the number of coordinates change from 3 to $D$. In this way, he sustained to prove space to be three-dimensional. Actually this is not the case. An epistemological criticism of this attempt to prove that space has 3 dimensions was made in Ref.~\cite{Moreira}.

This argument was also applied to the description of atomic orbits generated by a Coulomb-like potential, which is formally equivalent to the Newtonian one. In this case, it was showed that there is no bound state for hydrogen atom when $D\geq 5$.

Although Ehrenfest's approach to the problem of dimensionality, as Kant's conjecture, did not comply to prove what was originally proposed, proved to be a fertile idea. Frank Tangherlini, for example, in 1963, was the first to formally treat the problem of the hydrogen atom from the point of view of Schr\"{o}dinger equation \cite{Tangherlini}. His article inspired several others with nuances of difference~\cite{Mostepanenko}-\cite{Dirac}.

Many other physical phenomena were used in attempt to disclose the threefold nature of space. One can cite some examples of physical phenomena that depend on the number of dimensions of space in which they take place: neutron diffraction \cite{Moreira}, the Casimir effect \cite{Casimir-1,Casimir-2}, the stellar spectrum \cite{Danon} and the cosmic background radiation \cite{Cobe}. In the totality of these works -- as in all others dealing with the problem of space dimensionality -- \textit{time} is assumed to have just one dimension. Thus, even though it is known that a particular physical event in fact occurs in \textit{space-time}, one is actually discussing and imposing limits or constraints on the dimensionality of space alone, as is done, for example, in Ref.~\cite{Muller}. This is, however, a limited strategy to address the problem.

In fact, this limitation should not cause astonishment. The reason is that any discussion of the dimensionality of space or time, following Ehrenfest's general idea, \textit{i.e.}, searching for the singularities that the physical laws may present in relation to a particular number of dimensions, always strikes at an important point: the very fact that such laws are always theoretically or empirically determined without any kind of \textit{a priori} questioning about the dimensionality of space or time. Indeed, all natural law have been conceived admitting space to be three-dimensional and time to have just one dimension. It is as if these constraints were a fact of Nature, an unquestionable truth. This criticism is attenuated in the light of a new proposal recently published~\cite{AJP}.

In the case of time, this kind of prejudice seems to be even more ingrained in the scientific community, as suggested, for example, by the reference made earlier to Whithrow.

The sensitive experiences of temporal ordering and that time has one dimension seem so intertwined that, in fact, the literature on the problem of temporal dimensionality is incomparably small than dealing with the analogous characteristic of space. one can even query whether the very concept of causality on has inherited does not depend on having such relations as true.

So, in what experimental facts or in what other basic concepts should one who is interested in justifying that time is unidimensional or even prove that it can be multidimensional, be based upon? This theme should be treated elsewhere.

\section{The new background of non-Euclidean geometries}\label{non-euclidean}

In 1795, the Postulate number 5 of Euclid (Book I) was enounced by the English mathematician John Playfair as follow: Through a given point only one parallel can be drawn to a given straight line. This is known as the \textit{Parallel Postulate}~[39, v.~1, p.~220].

The Parallel Postulate started to be criticized by the German mathematician and physicist Johann Carl Friedrich Gauss -- who invented the concept of curvature --, in the last decade of the 18th century, when he tried to demonstrate it by using Euclidean Geometry. In effect, in 1792, when he was fifteen years old, he wrote a letter to his friend the German astronomer Heinrich Christian Schumacher, in which he discussed the possibility of having a Logical Geometry where the Parallel Postulate did not hold. In 1794, he conceived a new Geometry for which the area of a quadrangular figure should be proportional to the difference between $360^\circ$ and the sum of its internal angles. Later, in 1799, Gauss wrote a letter to his friend and Hungarian mathematician Wolfgang Farkas Bolyai saying that he had tried, without success, to deduce the Parallel Postulate from other postulates of Euclidean Geometry~\cite{Kaku}.

During the 19th century, Gauss continued the discussion with friends on the plausibility of the existence of a Non-Euclidean Geometry. So, around 1813, he developed what he initially called Anti-Euclidean Geometry, then Astral Geometry and, finally, Non-Euclidean Geometry. He was so convinced about the existence of this new Geometry that he wrote a letter, in 1817, to his friend and German astronomer and physician Heinrich Wilhelm Matth\"{a}us Olbers, stressing the physical necessity of such a Geometry as follow~[12, p.~871]:

\begin{quotation}
\noindent\baselineskip=10pt {\small
\noindent I am becoming more and more convinced that the [physical] necessity of our [Euclidean] geometry cannot be proved, at least not by human reason nor for human reason. Perhaps in another life we will be able to obtain insight into the nature of space, which is now unattainable. Until then we must place geometry not in the same class with arithmetic, which is purely a priori, but with mechanics.}
\end{quotation}

Seven years later, in 1824, answering a letter from the German mathematician Franz Adolf Taurinus talking about a demonstration he did that the sum of the internal angles of a triangle cannot be neither greater nor smaller than 180$^{\circ}$, Gauss told him that there was not geometrical rigor in that demonstration because, in spite of the fact that the ``metaphysicists'' consider the Euclidean Geometry as the truth, this Geometry is incomplete. The ``metaphysicists'' quoted by Gauss were the followers of Kant, who wrote, in 1781, in his \textit{Kritik der reinen Vernunft}~\cite{Kant-02}, more precisely in its first chapter entitled Transcendental Doctrine of Elements what follows: a) Space is not a conception which has been derived from outward experiences; b) Space then is a necessary representation \textit{a priori}, which serves for the foundation of all external intuitions; c) Space is represented as an infinite given quantity; d) Space has only three dimensions; d) (...) possibility of geometry, as a synthetic science \textit{a priori}, becomes comprehensible~\cite{Kant-03}.

While we owe to Gauss the discovery of Non-Euclidean Geometry, he did not have the courage to publish his discoveries. Indeed, in a letter sent to a German friend and astronomer Friedrich Wilhelm Bessel, in 1829, Gauss affirm that he probably would never publish his findings in this subject because he feared ridicule, or, as he put it, he feared the clamor of the Boetians, a figurative reference to a dull-witted Greek tribe~[12, p.~871].

In his research on the existence of a Non-Euclidean Geometry, Gauss figured out hypothetical ``worms'' that could live exclusively in a bi-dimensional surface, as other ``beings'' could be able to live in spaces of four or more dimensions~\cite{Silvester}. It is interesting to mention that, trying to verify his theory, Gauss and his assistants measured the angles of a triangle formed by the peaks of three mountains, Brocken, Hohehagen and Inselsberg, which belong to the Harz Mountais, in Germany. The distance between two of them were 69,85 and 197~km, respectively. The sum of the internal angles of this triangle was $180^\circ$ and 14'',85. This result frustrated Gauss since the error were within the errors associated to the instruments he used to measure the angles~\cite{Kline,Kaku}.

Independently of Gauss, the mathematicians, the Russian Nikolay Ivanovich Lobachevski and the Hungarian J\'{a}nos Bolyai (son of Wolfgang), in 1832, demonstrated the existence of triangles which sum of the internal angles are less than 180$^{\circ}$~\cite{Lobachevsky,Bolyai}.

The German mathematician Georg Friedrich Bernhard Riemann, after the presentation of his \textit{Doktoratsschrift}, in December 1851, in G\"{o}ttingen University, about the Fourier series and what is now know as Riemannian surfaces, started to prepare himself to become \textit{Privatdozent} of this same University. So, at the end of 1853, he presented his \textit{Habilitationsschrif} together with three topics for the \textit{Habilitationsvortrag}. For his surprise, Gauss choose the third topic entitled ``\"{U}ber die Hypothesen, welche der Geometrie zu Grunden liegen'' (``On the Hypothesis that are on the Base of Geometry''), where he demonstrated the existence of triangles of which the sum of its internal angles could be greater than 180$^{\circ}$. This topic was timidly presented by Riemann in June of 1854, but it provoked a deep impact on Gauss, because it was a concrete expression of his previous ideas about a Non-Euclidean Geometry (today, Riemannian Geometry) that he was afraid to publish, as previously mentioned. Riemann's metrical approach to Geometry and his interest in the problem of congruence also gave rise to another type of non-Euclidean geometry. We are talking about a new geometry that cames out not by the rejection of parallel axioms, but rather by its irregular curvature.

It is important to remember that those geometries, today generically know as Non-Euclidean Geometries~\cite{Beltrami,Loria} influenced physical thought in 19th century~\cite{Martins}. They are consequences of the observation that the relaxation of the Parallel Postulate could give rise to two new interpretations. One, in the Hyperbolic Geometry of Bolyai-Lobachevsky~\cite{Milnor}, for which, from a point outside a line, an infinite number of parallels can be drawn, and the second, in the Spherical Geometry of Riemann, where from a point outside a line, no parallel can be drawn to it~\cite{Kline,Bonola}.

Two specific areas of philosophical debate were the initial source of a \textit{sui generis} public interest in the non-Euclidean geometries and in the geometry in higher dimensions: the nature of the geometric axioms and the structure of our space~\cite{Henderson}. As time went on, a expressive interest of the general public fell on the nature of the space and the number of its dimensions. A historical record of this fact can be found in the accurate bibliography prepared by Duncan Sommerville, a Scottish mathematician and astronomer~\cite{Sommerville}. This history is well documented in the interesting book of Linda Henderson, historian of art~\cite{Henderson}. According to her, everything started with a movement to popularize $n$-dimensional spaces and non-Euclidean geometries in the second half of the 19th century. A whole literature was developed~\cite{Sommerville} around philosophical and mystical implications in relation to spaces of larger dimensions, easily accessible to a public of non-specialists; in particular, about the imagination of a fourth dimension, long before Minkowski's work and Einstein's Special Relativity and the Cubism. The popularization of these ideas contributed, as carefully analyzed in Ref.~\cite{Henderson}, to a revolution in Modern Art and, in particular, was fundamental to the Cubism, an artistic movement contemporary to Einstein's Special Relativity, where also use was made of non-Euclidean geometry, namely Minkowski's space-time.

\section{The fourth dimension as time-like component of the new space-time concept in Physics}\label{spacetime}

As shown in Ref.~\cite{fourth-dimens}, it was Riemann who generalized the concept of Geometries, by introducing the definition of metric, that defines how one can calculate the distance between two points, given by (in nowadays notation)
$$\mbox{d} s^2 = \sum_{i,j} = g_{ij} \mbox{d} x^i \mbox{d} x^j; \qquad (i,j = 1,2,3)$$
where $g_{ij}$  is the metric tensor of Riemann. In the case of flat spaces and rectilinear coordinates ($x,y,z$),
$$g_{ij} = (e_i, e_j) = \delta_{ij}$$
where $\delta_{ij}$ is the Kronecker delta, $e_i\, (i=1,2,3)$ are the vector-basis of a particular coordinate system and the notation $(e_i, e_j)$ means the scalar product between the two vectors.

Thus, the distance can be written as
$$\mbox{d} s^2 = \sum_{i,j} = \delta_{ij} \mbox{d} x^i \mbox{d} x^j\, = \, \mbox{d} x^2 + \mbox{d} y^2 + \mbox{d} z^2 $$
\noindent known as the Euclidean metric. This definition is straightforwardly extended to higher $n$-dimen\-sional spaces just doing $i,j \rightarrow \mu,\nu = 1,2,3, \cdots n$.

The Riemann work about Non-Euclidean Geometry (which easily allows the existence of more dimensions than the usual three), was soon recognized and flourish in all Europe, with eminent scientists propagating his ideas to the general public. For example, the German physicist and physiologist Hermann Ludwig Ferdinand von Helmholtz considered Gauss’ worms leaving now in a Riemannian surface (on a sphere). However, in his book entitled \textit{Popular Lectures of Scientific Subjects}, published in 1881, he warned that it is impossible to represent (to visualize) the fourth dimension, because (...) such a representation is so impossible how it should be a color representation for someone born blind~[36, p.~29].

From now on, let us summarize the route of the assimilation of such ideas in Physics.

The success of Newtonian mechanical World view will be put to the test, at first, by the study of heat made by the French mathematician and physicist Jean-Baptiste Joseph Fourier.

The propagation of heat will be described by a partial differential equation and no longer by an ordinary differential equation, as in the case of Newtonian mechanics. It is the beginning of valuing the \textit{causa formalis} over the \textit{causa efficiens} as the basis of the causal explanatory system in Physics, intrinsic to Newton's system~\cite{Causa_art}. It is the beginning of the description of Physics by Field Theories~\cite{Bachelard}. Later, in the second half of 19th century, also electromagnetism will reaffirm this trend~\cite{Oguri}.

The discovery of electromagnetic waves by the German physicist Heinrich Rudolf Hertz will give Maxwell's theory a new status. Nevertheless, Maxwell theory remains a phenomenological theory not able to predict, for example, the interaction of light with matter. One of the first attempts to develop a classical interpretive theory capable of explaining the interactions of electromagnetic fields with matter dates from 1895 and is due to the Dutch physicist Hendrik Antoon Lorentz, who combines the Electromagnetism and Classical Mechanics with an atomistic model of matter, the so-called Drude-Lorentz model,\footnote{\,The model according to which the physical world would be composed of ponderable matter, electrically charged mobile particles and ether, such that electromagnetic and optical phenomena would be based on the position and movement of these particles.} and initially develops a Newtonian Classical Electrodynamics, known as Lorentz Electrodynamics.

Soon after the electron discovery, this elementary particle gains a prominent place in theoretical physics~\cite{Buch}. In fact, as we have already mentioned, Lorentz will dedicate himself to include the interaction of this particle with the electromagnetic fields. As is well known, Lorentz Electrodynamics, despite some initial success, failed in correctly describe such kind of interaction. This problem will be solved just with the advent of Quantum Electrodynamics~\cite{Estranha}. From a conceptual point of view, Einstein attributed the weakness of Lorentz theory to the fact that it tried to determine the interaction phenomena by a combination of partial differential equations and total differential equations, a procedure that, in his opinion, is obviously not natural.

In 1888, the English Mathematician Oliver Heaviside showed that the electric field ($\vec E$) of a moving electric charge (with velocity $v$) differ from that ($\vec E_\circ$) of a stationary charge as indicated below~\cite{Heav}:
$$\vec E_\circ = \frac{kq}{r^2}\, \hat r \quad \Rightarrow \quad \vec E = \frac{kq}{r^2}\,\gamma\, \left[\frac{1 - \beta^2}{1 - \beta^2 \sin^2\theta} \right]^{3/2}\, \hat r$$
where $\beta = v/c$ and $\gamma = (1-\beta^2)^{-1/2}$. So, we can see that, in the direction of motion ($\theta =0$), the electric field behaves like
$$\vec E_\parallel = \frac{1}{\gamma^2}\frac{kq}{r^2} \hat r$$
Therefore, this result was interpreted by Heaviside as a contraction of the electrostatic field.

This result was published in 1889 and it was discussed by Heaviside, the British physicist Oliver Lodge and the Irish physicist George FitzGerald~\cite{Gray}. Inspired on this result, FitzGerald proposed that the objects contract along their line of flight. Independently, Lorentz came to the same idea in 1892 (see footnote in Ref.~\cite{Lorentz-1895}). This is the origin of Lorentz-FitzGerald contraction, involving the $\gamma$ factor.

Pre-Minkowskian applications of non-Euclidean geometry in Physics weren't many and they were reviewed in Ref.~\cite{Walter}.

Now, we would like to stress that, although Lorentz demonstrated, in 1904, that time is related to tri-dimensional space through the relations known as Lorentz Transformations (LT)~\cite{Lorentz-orig}, it was only the Russian-German mathematician Hermann Minkowski who understood~\cite{Minkowski} that the LT represent a kind of rotation in a 4-dimensional flat space having coordinates  ($x_1,x_2,x_3,x_4$), with a metric (measurement of the distance between two points in this space) defined by:

$$\mbox{d} s^2 = \sum_{\mu,\nu}^4 g^{\mu \nu} \mbox{d} x_\mu \mbox{d} x_\nu = \, \mbox{d} x_1^2 + \mbox{d} x_2^2 + \mbox{d} x_3^2 + \mbox{d} x_4^2$$

\noindent where $g^{\mu \nu} = \delta^{\mu \nu}$ is the four-dimensional Kronecker delta, $x_1=x$, $x_2=y$, $x_3=z$, $x_4= ict$, and $i =\sqrt{-1}$.

This expression is known as the Minkowskian metric, or pseudo-Euclidean metric, due to the fact that it can be negative. Note that, to avoid the use of $\sqrt{-1}$, the mathematicians defined a signature for  $g_{\mu\nu}$, such that the indices $\mu$ and $\nu$  can assume the values 1, 2, 3, 4 $(+, +, +, -)$ with $x^4 = ct$, or 0, 1, 2, 3 $(+, -, -, -)$ with $x^0 = ct$, where $\pm$ means $\pm 1$ only on the main diagonal of the metric tensor~\cite{Caruso-02}.

In his seminal paper of 1905 about the Electrodynamics of Moving Bodies, Einstein were able to derive LT without having to resort to ether by postulating the constancy of light velocity in the vacuum, \textit{i.e.}, assuming it does not depend on the velocity of the moving body~\cite{Einstein,Miller}.

For Lorentz, the local time ($t^\prime$) introduced in the coordinate transformations between inertial references, would be just an auxiliary parameter necessary to maintain the invariance of the laws of Electromagnetism, as stated at the end of the second edition of his \textit{Theory of Electrons}~[66, p.~321]:

\begin{quotation}
\noindent\baselineskip=10pt {\small
\noindent If I had to write the last chapter now, I should certainly have given a more prominent place to Einstein's theory of relativity (...) by which the theory of electromagnetic phenomena in moving systems gains a simplicity that I had not been able to attain. The chief cause of my failure was my clinging to the idea that the variable $t$ alone can be considered as the true time and that my local time $t^\prime$ must be regarded as no more than an auxiliary mathematical quantity. In Einstein's theory, on the contrary, $t^\prime$ plays the same part as $t$; if we want to describe phenomena in terms of $x^\prime,y^\prime, z^\prime, t^\prime$ we must work with these variables exactly as we could do with $x,y,z,t$.}
\end{quotation}

On the other hand, the conception and interpreta\-tion of Lorentz's trans\-for\-mations as a geometric transformation in a pseudo-Euclidean space of dimension 4, for Minkowski, was only possible thanks to Einstein's assertion, as quoted in a meeting of scientists in 1908, in Cologne~[67, p.~82]:
\begin{quotation}
\noindent\baselineskip=10pt {\small But the credit of first recognizing clearly that the time of the one electron is just as good as the time of the other, that $t$ and $t^\prime$ are to be treated identically, belongs to A. Einstein.}
\end{quotation}

However, another point made clear by Einstein is that for him the introduction of time as a fourth explicit coordinate in the transformations of inertial reference systems derives from the principle of relativity. In his words~[68, p.~365]:
\begin{quotation}
\noindent\baselineskip=10pt {\small It is a widespread error that special theory of relativity is supposed to have, to a certain extent, first discovered, or at any rate, newly introduced, the four-dimensionality of the physical continuum. This, of course, is not the case. Classical mechanics, too, is based on the four-dimensional continuum of space and time. But in the four-dimensional continuum of classical physics the subspaces with constant time value have an absolute reality, independent of the choice of the reference system. Because of this [fact], the four-dimensional continuum falls naturally into a three-dimensional and a one-dimensional (time), so that the four-dimensional point of view does not force itself upon one as \textit{necessary}. The special theory of relativity, on the other hand, creates a formal dependence between the way in which the spatial coordinates, on the other hand, and the temporal coordinates, on the other, have to enter into natural laws.}
\end{quotation}

\section{The role of metric space in the problem of space-time dimensionality}\label{role}

It is important to highlight the fact that a significant part of the arguments justifying the dimensionality of space depends on the existence of a metric space \cite{Moreira}. This fact refers to the notion of distance in a $n$ dimensional manifold, which is traditionally based on the homogeneous quadratic differential form
$$ \mbox{d} s^2 = g_{\mu\nu} \mbox{d}x^\mu \mbox{d}x^\nu$$

\noindent in which the $\mu$ and $\nu$ indexes assume the values $0,1,2,3,\cdots (n-1)$. This formula is ultimately an arbitrary choice, since, in fact, there are no logical arguments excluding \textit{a priori} other forms of the type $\mbox{d} s^4$, $\mbox{d} s^6$, $\mbox{d} s^8\cdots$.  At this point it is important to remember that, in 1920, Ehrenfest presented the conjecture that the exponent 2 of the quadratic form in the previous equation for the line element could be related to the dimensionality of space \cite{Ehrenfest1} but, to the best of our knowledge, such a conjecture has not yet been demonstrated. Some consequences, such as a possible relation of this conjecture with Fermat's theorem, is discussed in Ref.~\cite{Moreira}. Moreover, the fact that many of the fundamental equations of physics involve second-order spatial derivatives (Newton's equation, d'Alembert's wave equation, Schr\"{o}dinger's equation \textit{etc}.) may also be related to three-dimensionality of space. We will return to this point later.
	
{As already mentioned,} it was the development of non-Euclidean geometries in the nineteenth century \cite{Jammer} that allowed, well before the Theory of Relativity, the first speculations about a fourth dimension and what it would be. Among them we can quote that of Hinton \cite{Hinton-1,Hinton-2}.

Other aspects of the problem of reality (or not) of a fourth dimension, related to perception and philosophy, were treated by Whitrow \cite{Whitrow-1}. However, from the point of view of physics, it is the metric of Minkowski's geometry that can be easily generalized to any number of spatial and temporal dimensions. For $\mu$ and $\nu$ ranging from 0 to 3 ($n = 4$), we have the metric
$$ g^{\mu \nu} = \left(
                    \begin{array}{cccc}
                      + & 0 & 0 & 0 \\
                      0 & - & 0 & 0 \\
                      0 & 0 & - & 0 \\
                      0 & 0 & 0 & - \\
                    \end{array}
                  \right)$$

For any number of space-time dimensions $n$, the new array $g^{\mu\nu}$ will have dimensions $n\times n$. In his famous book \textit{The Mathematical Theory of Relativity}, Arthur Stanley Eddington ponders that such a choice (a sign $+$ and three $-$) specifies the world in a way that we could hardly have predicted from first principles \cite{Eddington-Rel}. Why does space-time have one and not another signature? It reminds us, then, of the English astrophysicist, without citing the reference, that Hermann Weyl expresses this ``special'' character by stating that space has dimensions \cite{Weyl}. However, a careful reading of his works shows that this would be the total number of space-time dimensions that ensures the scale invariance of Maxwell's Classical Electromagnetism, but with the time-dimensionality fixed at~1.

Back to Eddington's book, he examines another interesting question: whether the universe can change its geometry. In particular, one wonders if in some remote region of space or time one could have a metric of the type
$$ g^{\mu \nu} = \left(
                    \begin{array}{cccc}
                      - & 0 & 0 & 0 \\
                      0 & - & 0 & 0 \\
                      0 & 0 & - & 0 \\
                      0 & 0 & 0 & - \\
                    \end{array}
                  \right)$$

Its answer is negative and the argument is that, if such a region exists, it must be separated by a surface of the region in which the metric signature is $(+, -, -, -)$, so that, for a side of the separation surface, there is
$$ \mbox{d} s^2 = c_1^2\mbox{d} t^2 - \mbox{d} x^2 - \mbox{d} y^2 - \mbox{d} z^2$$
while, on the other side,
$$ \mbox{d} s^2 = - c_2^2\mbox{d} t^2 - \mbox{d} x^2 - \mbox{d} y^2 - \mbox{d} z^2$$

The transition, in this case, could only take place through a surface on which
$$ \mbox{d} s^2 = 0 \mbox{d} t^2 - \mbox{d} x^2 - \mbox{d} y^2 - \mbox{d} z^2$$

Therefore, the (fundamental) velocity $c$ of light would be null~\cite{Eddington-Rel},
\begin{quotation}
\noindent\baselineskip=10pt{\small{Nothing could move on the surface of separation between the two regions and no influence can pass from one side to another. The supposed ulterior region is not in any space-time relationship with our universe -- which is a somewhat pedantic way of saying that it does not exist.} }
\end{quotation}

It is now known that in the classical theories of Gravitation there can be no local changes in the space-time topology without considering quantum fluctuations \cite{Ivano}. The case of a hypothetical world of dimensions $2+2$ is also briefly discussed by the author. The possibility of a universe in which time can be two-dimensional is still dealt with in another posthumously published book by Eddington \cite{Eddington-Fund}.
	
In 1970, Dorling developed an essentially kinematic argument showing that \cite{Dorling}
\begin{quotation}
\noindent\baselineskip=10pt{\small{the extreme [maximum] property of time-type geodesics in an ordinary space-time is a necessary condition for the existence of stable particles. This maximum property would fail if time were multidimensional.}}
\end{quotation}

According to this author, for a multidimensional time, the proton and the electron would not be stable. Not even the photon! In addition, it proposes that objections that occur at speeds greater than that of light and for a multidimensional time may be related. This opens up a possibility of studying \textit{tachyons} in universes with a greater number of time-type coordinates. This is because the only essential difference between time and space (and between the corresponding time-type or space type geodesics) in a Minkowski geometry is the difference in their dimensionality.

Also worthy of note is the work of Mirman \cite{Mirman}, in which he defends the thesis that the space-time signature seems to be related to the measurement process and, if there is more than one dimension of the time type, the extra dimensions do not would be observable. The question of measurement seems to us to be a central point in this whole discussion of the number of dimensions, whether of space or of time.

Any measurement process depends not only on the definition of an observer but also on some physical laws. Often what is done to discuss the problem of the dimensionality of space is to generalize the functional form of a differential equation that describes a law of physics in a space $R^3$ (which -- it is always necessary to remember -- has been established without any kind of questioning about the three-dimensionality of space) for a space $R^n$, but keeping the order of the differential equation. Thus, when discussing planetary stability based on Newtonian gravitation in spaces of arbitrary dimensions what is done is to generalize the Poisson equation as follows:
$$\nabla^2_{(3)} \phi = \frac{\partial^2}{\partial x_1^2} + \frac{\partial^2}{\partial x_2^2} + \frac{\partial^2}{\partial x_3^2} = 4\pi \rho \ \ \Rightarrow \ \ \nabla^2_{(n)} \phi = \frac{\partial^2}{\partial x_1^2} + \frac{\partial^2}{\partial x_2^2} + \cdots + \frac{\partial^2}{\partial x_n^2} = 4 \pi \rho $$

From this we find the general solution of a generalized equation, and assuming a hypothesis (justified only by arguments of anthropic nature \cite{Methane}) that it similarly describes the same physical phenomenon of the case $n=3$, we discuss the mechanical stability of this new solution. The epistemological limitations of this method were extensively discussed in Ref.~\cite{Moreira} (See also Ref.~\cite{AJP}).
	
Thus, at this point, the aim of the paper is achieved.

\section{Giambiagi and Bolini on multidimensional time}\label{roots}

Since this paper is not aimed to review what has been done on multidimensional time, the scope of this small Section is just highlights the original and inspiring contribution of Giambiagi and Bollini for discussing space and time dimensionality, following Ehrenfest's Ansatz.

At this point we would like to recall some works by the dear and late friend Juan Jos\'{e} Giambiagi, who, with Guido Bollini, developed in 1972 the famous method of dimensional regularization \cite{RD-1,RD-2}, admitting that the dimensionality of space-time is a real number given by $\nu = 3 + 1 - \epsilon$.

J.J. (as he was known among friends) worked with different collaborators on the problem of dimensionality of space and time in a very open way without any kind of prejudice \cite{JJ-1}-\cite{JJ-12}. In these seminal works of 1972, Bollini \& Giambiagi showed for the first time (as far as we know) that a small fluctuation imposed \textit{ad hoc} to the dimensionality of space-time is at the basis of a method capable of controlling divergences arising in calculating certain physical quantities in gauge field theories. Therefore, they have shown that the dynamics in a field theory can also depend crucially on the number of space-time dimensions.
	
In the following articles, in very general terms, Giambiagi and collaborators place particular emphasis on the study of the generalized d'Alembert wave equation and its relation to the Huygens principle. What is important for the scope of this essay is to draw attention to the fact that they do so in a much more general way than the generalization of the above-mentioned Poisson equation, with increasing sophistication for each article, even allowing new potencies for the d'Alembert operator, $\Box$. The fact that the properties of the wave equations strongly depend on the spatial dimensions is not new and had already been noted by Ehrenfest \cite{Ehrenfest1,Ehrenfest2}, Henri Poincar\'{e} \cite{Poincare} and Jacques Hadamard \cite{Hadamard}. The argument that spatially three-dimensional worlds seem to have a unique and very peculiar combination of properties that guarantee the processing and propagation of signals via electromagnetic phenomena can be found in \cite{Barrow}, but it is worth emphasising that this argument is built on the general perception that time has one dimension.
	
Motivated by new developments in Gravitation and Supersymmetry Theories, Giambiagi seeks to free himself from this ``prejudice'' and will study, in several of the articles already cited here, from the point of view of Mathematical Physics, solutions for different dimensions of equations involving operators $\Box$, $\Box^{1/2}$, $\Box^2$, $\Box^3$, $\Box^\alpha$, for a temporal coordinate, initially \cite{JJ-8}, and then in a space-time with $(p+q)$ dimensions \cite{JJ-10,JJ-11,JJ-12}, where
$$\Box = \frac{\partial^2}{\partial t_1^2} + \frac{\partial^2}{\partial t_2^2} + \cdots + \frac{\partial^2}{\partial t_q^2} - \frac{\partial^2}{\partial x_1^2} - \frac{\partial^2}{\partial x_2^2} - \cdots - \frac{\partial^2}{\partial x_p^2}$$

And the solutions analyzed are those that depend only on the variables
$$ t = \sqrt{t_1^2 + t_2^2 + \cdots + t_q^2} \qquad \mbox{e} \qquad r = \sqrt{x_1^2 + x_2^2 + \cdots + x_p^2}$$

It is easy to see that the epistemological nature of an eventual linkage of these results, showing that only the power $\alpha = 1$ of the $\Box$ operator and a four-dimensional space-time would ensure the propagation of electromagnetic waves without problems of loss of information and without reverberations~\cite{Barrow}, would be very different from the already known result. Other more recent contributions in this field of Mathematical Physics can be found in Refs.~\cite{Craig, Weinstein}.

\section{Concluding remarks}\label{conclusion}

All these works offer a substantial range of results that deserve to be analyzed from an epistemological point of view, and not only from the formal perspective, according to which the multiplicity of spatial and temporal dimensions is only a mathematical possibility to be explored and investigated. Perhaps this kind of inquiry into some of these results may shed light on how much our perception and the formal adoption of one-dimensional time on the one hand and the functional forms of physical laws on the other are intertwined. Or perhaps, {the scenario is intrinsically more limited, as expressed by Weinstein'a quotation: ``[theories with multiple spatial and temporal dimensions] serve to broaden our minds in the sense of what may be physically possible.''
\vspace*{0.5cm}

\renewcommand\refname{References}


\begin{thebibliography}{50}
\bibitem{Friedman}
W. Friedman. \textit{About Time: inventing the fourth dimension}. Cambridge, Massachusetts: The MIT Press (1990), p.~6.

\bibitem{Juliao}
J. N. Juli\~{a}o. Time and History by St.~Augustine. \textit{Veritas} \textbf{63} (2), pp.~408-435 (2018).

\bibitem{Hernandez}
W. A. Hernandez. St.~Augustine on Time. \textit{International Journal of Humanities and Social Science} \textbf{6} (6), pp.~37-40 (2016).

\bibitem{Augustine}
Augustine. \textit{The Confessions}. Ed.~M. P. Foley. Indianapolis: Hackett Publishing Company (2006).

\bibitem{Whitrow-1}
G. J. Whitrow. \textit{The Natural Philosophy of Time}. Oxford: University Press, second edition (1980), pp.~368-70.

\bibitem{Whitrow-2}
\textit{Idem}, p.~370.

\bibitem{Barnes}
J. Barnes, (Ed.): \textit{The Complete Works of Aristotle}. The revised Oxford Translation, Princeton: Princeton University Press, volume 1, p.~447 (1984).

\bibitem{Pauli}
W. Pauli. The influence of Archetypal Ideas on the Scientific Theories of Kepler. \textit{In} C. G.~Jung \& W. Pauli: \textit{The Interpretation of Nature and the Psyche}. Japan: Ishi Press International, 2012.

\bibitem{Jung}
C. G. Jung. \textit{O Homem e seus S\'{\i}mbolos}. Rio de Janeiro: Editora Nova Fronteira, p.~307 (1992).

\bibitem{fourth-dimens}
J. M. F. Bassalo, F. Caruso and V. Oguri. The Fourth Dimension: From its spatial nature in Euclidean geometry to a time-like component of non-Euclidean manifolds. \textit{Revista Brasileira de Ensino de F\'{\i}sica} \textbf{43} (2021) e20210034.

\bibitem{Pascal}
B. Pascal. \textit{{\OE}uvres}. Paris: Lefr\`{e}ve Libraian (1819), v.~5, p.~260-298.

\bibitem{Kline}
M. Kline. \textit{(Mathematical Thought from Ancient to Modern Times}. Oxford: Oxford University Press (1972).

\bibitem{Moebius}
A. F. M\"{o}bius. \textit{Der barycentrische Calcul, ein neues H\"{u}lfsmittel zur anaytischen Behandlung der Geometrie}. Leipzig: Barth (1827).

\bibitem{Manning}
H. P. Manning. \textit{Geometry of Four Dimensions}. New York: The Macmillan Company (1914).

\bibitem{Boyer}
C. B. Boyer. \textit{A History of Mathematics}. New York: John Wiley and Sons (1968).

\bibitem{Caruso-02}
F. Caruso. Ensaio sobre a Dimensionalidade do Tempo. \textit{Tempo Brasileiro} \textbf{195}, pp.~83-90 (2014).

\bibitem{Kant}
J. Handyside (Ed.). \textit{Kant's inaugural dissertation and the early writings on space}. Chicago: Open Court (1929), reprinted by Hyperion Press (1979),  pp.~11-15.

\bibitem{Epistemologia}
F. Caruso \& R. Moreira. Sull'influenza di Cartesio, Leibniz e Newton nel primo approccio di Kant al problema dello spazio e della sua dimensionalit\`{a} (in Italian). \textit{Epistemologia (Genova, Italia)} \textbf{21} (1998) 211-224.

\bibitem{Kant-Studien}
F. Caruso \& R. Moreira. ``On Kant's first insight into the problem of space dimensionality and its physical foundations''. \textit{Kant-Studien} \textbf{106} (2015) 547-560.

\bibitem{Britain}
G. G. Brittan Jr. \textit{Kant's Theory of Science}. Princeton: Princeton University Press (1978).

\bibitem{Postumum}
E. Kant. \textit{Opus Postumum -- passage des principes m\'{e}taphysiques de la science de la nature \`{a} la physique}. Translation, presentation and notes by F. Marty. Paris, Presses Universitaire de France (1986), p.~131.

\bibitem{Ehrenfest1}
P. Ehrenfest. In what way does it become manifest in the fundamental laws of physics that space has three dimensions? \textit{Royal Netherlands Academy of Arts and Sciences (KNAW)} \textbf{20} (1917) 200. Reprinted in M. J.~Klein (Ed.). {\it Paul Ehrenfest -- Collected Scientific Papers}. Amsterdam: North Holland Publ.~Co. (1959), p.~400.

\bibitem{Ehrenfest2}
P. Ehrenfest. Welche Rolle spielt die Dreidimensionalit\"{a}t des Raumes in den Grund\-gesetzen der Physik? \textit{Annalen der Physik} \textbf{61} (1920) 440.

\bibitem{Causa}
F. Caruso \& R. Moreira, \textit{Causa efficiens} versus \textit{causa formalis}: origens da discuss\~{a}o moderna sobre a dimensionalidade do espa\c{c}o (in Portuguese). \textit{Scientia (Unisinos)} \textbf{4} (1994) 43-64.

\bibitem{Moreira}
F. Caruso \& R. Moreira Xavier. On the Physical Problem of Spatial Dimensions: An Alternative Procedure to Stability Arguments. \textit{Fundamenta Scientiae} \textbf{8} (1988) 73.

\bibitem{Tangherlini}
F. R. Tangherlini. Schwarzschild Field in $n$ Dimensions and the Dimensionality of Space Problem. \textit{Nuovo Cimento} \textbf{27} (1963) 636-651.

\bibitem{Mostepanenko}
L. Gurevich and V. Mostepanenko. On the existence of atoms in $n$-dimensional space. \textit{Physics Letters A} \textbf{35} (1971) 201.

\bibitem{Supplee}
K. Andrew and J. Supplee. A hydrogenic atom in $d$-dimensions. \textit{American Journal of Physics} \textbf{58} (1990) 1177.

\bibitem{Giambiagi}
C. G. Bollini, J. J. Giambiagi \& J. S. Helman. Study of wave equations involving iterated Laplacian and potential $r^{-\beta}$ by the $1/N$ method. \textit{Chemical Physics Letters} \textbf{175} (1990) 130.

\bibitem{Shaqqor}
M. A. Shaqqor and S. M. AL-Jaber. A Confined Hydrogen Atom in Higher Space Dimensions. \textit{Internation Journal of Theoretical Physics} \textbf{48} (2009) 2462.

\bibitem{Caruso}
F. Caruso, J. Martins and V. Oguri. On the existence of hydrogen atoms in higher dimensional euclidean spaces. \textit{Physics Letters A} \textbf{377} (2013) 694.

\bibitem{Dirac}
F. Caruso, J. Martins, L. D. Perlingeiro \& V. Oguri. Does Dirac equation for a generalized Coulomb-like potential in $D+1$ dimensional flat space–time admit any solution for $D \geq 4$? \textit{Annals of Physics} \textbf{359} (2015) 73.

\bibitem{Casimir-1}
F. Caruso, N. P. Neto, B. F. Svaiter \& N. F. Svaiter. Attractive or repulsive nature of Casimir force in $D$-dimensional Mikowski spacetime. \textit{Physical Review D} \textbf{43} (1991) 1300-1306.

\bibitem{Casimir-2}
F. Caruso, R. De Paola \& N.F. Svaiter. Zero point energy of massless scalar field in the presence of soft and semihard boundary in $D$ dimensions. \textit{International Journal of Modern Physics A} \textbf{14} (1999) 2077-2089.

\bibitem{Danon}
F. Caruso \& R. Moreira. Space dimensionality: what can we learn from stellar spectra and from the M\"{o}ssbauer effect. \textit{In}: R. B. Scorzelli, I. Souza Azevedo \& E. Baggio Saitovitch (Eds.), \textit{Essays on Interdisciplinary Topics in Natural Sciences Memorabilia: Jacques A. Danon}, Gif-sur-Yvette/Singapore: \'{E}ditions Fronti\`{e}res, pp.~73-84 (1997).

\bibitem{Cobe}
F. Caruso \& V. Oguri. The Cosmic Microwave Background Spectrum and a Determination of Fractal Space Dimensionality. \textit{Astrophysical Journal} \textbf{694} (2009) 151-153.

\bibitem{Muller}
B. M\"{u}ller \& A. Sch\"{a}fer. Improved bounds on the dimension of space-time. \textit{Physical Review Letters} \textbf{56} (1986) 1215-1218.

\bibitem{AJP}
F. Caruso; V. Oguri \& F. Silveira. Still learning about space dimensionality: From the description of hydrogen atom by a generalized wave equation for dimensions $D \geq 3$. \textit{American Journal of Physics} \textbf{91} (2023) 153–158.

\bibitem{Euclid}
Euclid. \textit{The Thirteen books of Euclid's Elements}. New York: Dover (1956).

\bibitem{Kaku}
M. Kaku. \textit{Introduction to Superstrings and M-Theories}. Berlin: Springer-Verlag (1999).

\bibitem{Kant-02}
I. Kant. \textit{Kritik der reinen Vernunft}. English translation in \textit{Great Books of the Western World}, vol.~39. Chicago: Encyclop\ae{dia} Britannica, Inc. (1993).

\bibitem{Kant-03} Kant, I.: \textit{idem}, pp.~24-25.

\bibitem{Silvester}
J. J. Silvester. A Plea for the Mathematicians. \textit{Nature}, v.~1, p.~238 (1869).

\bibitem{Lobachevsky}
N. I. Lobachevsky. G\'{e}om\'{e}trie imaginaire. \textit{Journal f\"{u}r die Reine und Angewandte Mathematik}, v.~17, p.~295 (1837).

\bibitem{Bolyai}
J. Bolyai. \textit{Absoluten Geometrie}. B.G. Teubner, Leipzig (1832). Indeed, Bolyai first announced his discoveries in a 26 pages appendix to a book by his father, the Tentamen, published in (1831).

\bibitem{Beltrami}
E. Beltrami. Saggio di interpretazione della geometria non-euclidea. \textit{Giornale di Matematiche}, VI, pp.~284-322 (1868).

\bibitem{Loria}
G. Loria. \textit{Il passato e il presente delle principali teorie geometriche. Storia e bibliografia}, fourth edition. Padova: Cedam (1931).

\bibitem{Martins}
R. Martins. The influence of non-Euclidean geometries in 19th century physical thought (in Portuguese). \textit{Revista da Sociedade Brasileira de Hist\'{o}ria da Ci\^{e}ncia}, v.~13,
pp.~67-79 (1995).

\bibitem{Milnor}
J. Milnor. Hyperbolic Geometry: The first 150 Years. \textit{Bulletin of The American Mathematical Society} (New Series) v.~6, n.~ 1, pp.~9-24 (1982).

\bibitem{Bonola}
R. Bonola. \textit{Geometr\'{\i}as No Euclidianas}. Buenos Aires: Espasa-Calpe Argentina (1951).

\bibitem{Henderson}
L. D. Henderson. \textit{The Fourth Dimension and Non-Euclidean Geometry in Modern Art}. Princeton: New Jersey, Princeton University Press (1983).

\bibitem{Sommerville}
D. M. Y. Sommerville. \textit{Bibliography of Non-Euclidean Geometry including the Theory of Parallels, the Foundations of Geometry and Space of $n$ Dimensions}. St.~Martin's Lane, London: Harrison \& Sons (1911).

\bibitem{Causa_art}
F. Caruso \& R. Moreira Xavier. \textit{Causa efficiens versus causa formalis}: origens da discu\c{c}\~{a}o moderna sobre a dimensionalidade do espa\c{c}o f\'{\i}sico. \textit{Cadernos de Hist\'{o}ria e Filosofia da Ci\^{e}ncia}, S\'{e}rie 3, \textbf{4}, n.~2, p. 43-64, jul.-dez. (1994).

\bibitem{Bachelard}
R. Moreira Xavier. Bachelard e o livro do calor: o nascimento da F\'{\i}sica Matem\'{a}tica na \'{e}poca da rearticula\c{c}\~{a}o causal do Mundo. \textit{Revista Filos\'{o}fica Brasileira} \textbf{6}, p.~100-113 (1993).

\bibitem{Oguri}
F. Caruso \& V. Oguri. \textit{F\'{\i}sica Moderna: Origens Cl\'{a}ssicas e Fundamentos Qu\^{a}nticos}. Second edition, Rio de Janeiro: Editora LTC (2016).

\bibitem{Buch}
J. Z. Buchwald. \textit{From Maxwell to microphysics}. Chicago: Chicago University Press (1985).

\bibitem{Estranha}
F. Caruso \& V. Oguri. A estranha teoria da luz. \textit{Ci\^{e}ncia e Sociedade} \textbf{4}, n.~2, pp.~12-17 (2016).

\bibitem{Heav}
O. Heaviside. Electromagnetic waves, the propagation of potential, and the electromagnetic effects of a moving charge. Reproduced in \textit{Electrical Papers}, v.~2, pp.~490-499 (1894).

\bibitem{Gray}
J.J. Gray (Ed.). \textit{The Symbolic Universe: Geometry and Physics 1890-1930}. New York: Oxford University Press (1999).

\bibitem{Lorentz-1895}
H. A. Lorentz. Versuch einer Theorie der elektrischen und optischen Erscheinungen in betwegten K\"{o}rper, Leiden: Brill. Reproduced in \textit{Collected Papers}, v.~5, pp.~1-137.

\bibitem{Walter}
S. Walter. The non-Euclidean style of Minkowskian relativity. In \textit{The Symbolic Universe: Geometry and Physics 1890-1930}, \textit{op.~cit.}, pp.~91-127 (1999).

\bibitem{Lorentz-orig}
H. A. Lorentz. Electromagnetic phenomena in a system moving with any velocity smaller than that of light. \textit{Proceedings Acad. Sc. Amsterdam} \textbf{6}, p.~809 (1904).

\bibitem{Minkowski}
H. Minkowski. Die Grundgleichungen f\"{u}r die elektromagnetischen Vorg\"{a}nge in bewegten K\"{o}rpen. \textit{K\"{o}niglich Gesellschaft der Wissenschaften zu G\"{o}ttingen Nachrichten, Mathematisch-Physikalische Klasse}, p.~53 (1908).

\bibitem{Einstein}
A. Einstein. Zur Elektrodynamik bewegter K\"orper. \textit{Annalen der Physik}, Ser.~4, \textbf{17}, pp.~891-921 (1905). Portuguese Translation in J. Stachel (Org.): \textit{O ano miraculoso de Einstein: cinco artigos que mudaram a face da F\'{\i}sica}. Rio de Janeiro: Ed. UFRJ (2001).

\bibitem{Miller}
T. S. Miller. \textit{Albert Einstein's Special Theory of Relativity; Emergence (1905) and early interpretation (1905-1911)}. Massachusetts: Addison-Wesley (1981).

\bibitem{Lorentz}
H. A. Lorentz. \textit{The Theory of Electrons and its applications to the phenomena of light and radiant heat}. Leipzig: Teubner (1909).

\bibitem{Cologne}
H. Minkowski. Raum und Zeit. \textit{Physikalische Zeitschrift}, v.~20, pp.~104-111 (1909) or Space and Time. A Translation of an Address delivered at the 80th Assembly of German Natural Scientists and Physicians, at Cologne, 21 September (1908).

\bibitem{Jammer}
 M. Jammer. \textit{Concepts of Space: The History of Theories of Space in Physics}. New York: Dover (1993).

\bibitem{Hinton-1}
C. H. Hinton. \textit{What is the fourth dimension?}. London: Allen und Unwin (1887).

\bibitem{Hinton-2}
C. H. Hinton. \textit{Speculations on the Fourth Dimension: Selected Writings of Charles H. Hinton}, New York: Dover (1980).

\bibitem{Eddington-Rel}
A. S. Eddington. \textit{The Mathematical Theory of Relativity}. Cambridge: University Press (1923), p.~25.

\bibitem{Weyl}
H. Weyl. Gravitation und Elektrizit\"{a}t. \textit{Sitzungsberichte der K\"{o}niglich Preu{\ss}ischen Akademie der Wissenschaften zu Berlin}, p.~465-480 (1918). Eine neue Erweiterung der Relativit\"{a}tstheorie. \textit{Annalen der Physik} \textbf{59}, pp.~101-133 (1919). See also his \textit{Space, Time, Matter}. New York: Dover (1952), pp.~282-825.

\bibitem{Ivano}
See, for exemple, J. Martin, N. Pinto-Neto \& I. Dami\~{a}o Soares. Green functions for topology change. \textit{Journal of High Energy Physics} \textbf{3} (2005) 60, and references therein.

\bibitem{Eddington-Fund}
A. S. Eddington. \textit{Fundamental Theory}. Cambridge: University Press (1946), p.~126.

\bibitem{Dorling}
J. Dorling. The dimensionality of time. \textit{American Journal of Physics} \textbf{38} (1970) 539-540.

\bibitem{Mirman}
R. Mirman. Comments on the dimensionality of time. \textit{Foundations of Physics} \textbf{3} (1973) 321-333.

\bibitem{Methane}
F. Caruso. A note on space dimensionality constraints relied on anthropic arguments: Methane structure and the origin of life. \textit{In} M. S. D. Cattani; L. C. B. Crispino; M. O. C. Gomes \& A. F. S. Santoro (Eds.) \textit{Trends in Physics: Festschrift in homage to Prof. Jos\'{e} Maria Filardo Bassalo}, S\~{a}o Paulo: Livraria da F\'{\i}sica (2009), pp.~95-106.

\bibitem{RD-1}
C. G. Bollini \& J. J. Giambiagi. Dimensional Regularization: The Number of Dimensions as a Regularizing Parameter. \textit{Nuovo Cimento B} \textbf{12} (1972) 20-26.

\bibitem{RD-2}
C. G. Bollini \& J. J. Giambiagi. Lowest order `divergent' graphs in $\nu$-dimensional space. \textit{Physics Letters B} \textbf{40} (1972) 566-568.

\bibitem{JJ-1}
C. G. Bollini \& J. J. Giambiagi. Supersymmetric Klein-Gordon equation in $d$-dimensions. \textit{Physical Review D} \textbf{32} (1985) 3316-3318.

\bibitem{JJ-2}
C. G. Bollini \& J. J. Giambiagi. Lagrangian Procedures for Higher order field equation. \textit{Revista Brasileira de F\'{\i}sica} \textbf{17} (1987) 14-30.

\bibitem{JJ-3}
C. G. Bollini \& J. J. Giambiagi. Higher order equations of Motion. \textit{Revista Mexicana de F\'{\i}sica} \textbf{36} (1990) 23-29.

\bibitem{JJ-4}
C. G. Bollini \& J. J. Giambiagi. Huyghens' Principle in $(2n+1)$ Dimensions for Nonlocal Pseudodifferential Operator of the Type $\Box^\alpha$. \textit{Nuovo Cimento A} \textbf{104} (1991) 1841-1844.

\bibitem{JJ-5}
C. G. Bollini, J. J. Giambiagi \& O. Obreg\'{o}n. Are some physical theories related with a specific number of dimensions?. In A. Feinstein \& J. Ib\'{a}\~{n}ez (Eds.), \textit{Recent Developments in Gravitation} (Proceedings of Spanish Conference on Gravitation), Singapore: World Scientific (1992), p.~103.

\bibitem{JJ-6}
C. G. Bollini \& J. J. Giambiagi. Criteria to Fix the Dimensionality Corresponding to Some Higher Derivative Lagrangians. \textit{Modern Physics Letters A} \textbf{7} (1992) 593-599.

\bibitem{JJ-8}
C. G. Bollini \& J. J. Giambiagi. ``Arbitrary Powers of d'Alembertians and the Huygens' Principle'', \textit{Journal of Mathematical Physics} \textbf{34} (1993) 610-621.

\bibitem{JJ-9}
J. J. Giambiagi. ``Relations Among Solutions for Wave and Klein-Gordon Equations for Different Dimensions'', \textit{Nuovo Cimento B} \textbf{109} (1994) 635-644.

\bibitem{JJ-10}
W. Bietenholz \& J. J. Giambiagi. ``Solutions of the Spherically Symmetric Wave Equation in $p+q$  dimensions'', \textit{Journal of Mathematical Physics} \textbf{36} (1995) 383-397.

\bibitem{JJ-11}
C. G. Bollini, J. J. Giambiagi, J. Benitez \& O. Obreg\'{o}n. ``Which is the Dimension of Space if Huygens' Principle and Newtonian Potential are Simultaneously Satisfied?'', \textit{Revista Mexicana de F\'{\i}sica} \textbf{39}, suplemento n.~1, (1993) S1-S6.

\bibitem{JJ-12}
J. J. Giambiagi. ``Wave Equations with multiple times: Classical and Quantum Solutions'', \textit{preprint} CBPF-NF-055 (1995).

\bibitem{Poincare}
H. Poincar\'{e}. \textit{Derni\`{e}res Pens\'{e}es}. Paris: Flammarion (1917).

\bibitem{Hadamard}
J. Hadamard. \textit{Lectures on Cauchy's problem in linear partial differential equations}. New Haven: Yale University Press (1923).

\bibitem{Barrow}
J. D. Barrow, Dimensionality. \textit{Philosophical Transactions of the Royal Society of London A} \textbf{310} (1983) 337-346.

\bibitem{Craig}
W. Craig \& S. Weinstein. On determinism and well-posedness in multiple time dimensions. \textit{Proceedings of the Royal Society A} \textbf{465} (2009) 3023–3046.

\bibitem{Weinstein}
S. Weinstein, Multiple time dimensions. \url{arXiv:0812.3869v1} (2008).


\end{thebibliography}
\end{document}